\documentclass[sigconf,authorversion,nonacm]{acmart}

\AtBeginDocument{%
  \providecommand\BibTeX{{%
    \normalfont B\kern-0.5em{\scshape i\kern-0.25em b}\kern-0.8em\TeX}}}

 \acmConference[ApPLIED 2022]{Part of PODC 2022}{July 25,
   2022}{Sorrento, Italy}
\usepackage[normalem]{ulem}
\usepackage[noend]{algpseudocode}
\usepackage{subfig}

\begin{document}

\title{Towards an Approximation-Aware Computational Workflow Framework for Accelerating Large-Scale Discovery Tasks}
\author{Michael A. Johnston}
\affiliation{%
  \institution{IBM Research Europe - Dublin}
  \streetaddress{Damastown}
  \city{Dublin}
  \country{Ireland}
}
\email{michaelj@ie.ibm.com}

\author{Vassilis Vassiliadis}
\affiliation{%
  \institution{IBM Research Europe - Dublin}
  \streetaddress{Damastown}
  \city{Dublin}
  \country{Ireland}
}
\email{vassilis.vassiliadis@ibm.com}


\begin{abstract}
The use of approximation is fundamental in computational science.
Almost all computational methods adopt approximations in some form in order to obtain a favourable cost/accuracy trade-off and there are usually many approximations that could be used. 
As a result, when a researcher wishes to measure a property of a system with a computational technique, they are faced with an array of options.
Current computational workflow frameworks focus on helping researchers automate a sequence of steps on a particular platform.
The aim is often to obtain a computational measurement of a property.
However these frameworks are unaware that there may be a large number of ways to do so. As such, they cannot support researchers in making these choices during development or at execution-time.  

We argue that computational workflow frameworks should be designed to be \textit{approximation-aware} - that is, support the fact that a given workflow description represents a task that \textit{could} be performed in different ways. 
This is key to unlocking the potential of computational workflows to accelerate discovery tasks, particularly those involving searches of large entity spaces.  
It will enable efficiently obtaining measurements of entity properties, given a set of constraints, by directly leveraging the space of choices available. 
In this paper we describe the basic functions that an approximation-aware workflow framework should provide, how those functions can be realized in practice, and illustrate some of the powerful capabilities it would enable, including approximate memoization, surrogate model support, and automated workflow composition.
\end{abstract}


\maketitle


\section{Introduction}

Computational workflows are popular research accelerators for a broad range of scientific domains~\cite{taylor2007workflows, da2017characterization}. 
They are often described as graphs where nodes represent tasks and edges define data dependencies. 
Tasks range from lightweight applications to resource-intensive simulations with long execution times. 
Computational workflow frameworks focus on encoding and deploying these complex graphs of tools and services, tackling topics like step-automation, automating data-transfer, and automating the collection of provenance and other metadata for reproducibility \cite{kubeflow, argo, tekton, guixwl, sevenbridges, guixwl, rayframework, rayworkflows, di2017maestro, deelman2015pegasus}. 

Computational workflows often comprise initialization, computation, and analysis steps that can process large volumes of data~\cite{kelling2009data,draxl2018nomad, smith2020molssi, molssidriver2021, mcdonagh2021can}. 
Frequently, the purpose of these steps is to  measure characteristics of input systems.
Hence such computational workflows can be thought of as \textit{virtual-experiments}. 
In this mode they are critical for discovery tasks e.g. materials discovery, that involve searching large spaces of entities for systems with desired properties. 

In our experience, cost drives developments in computational science to a greater extent than other computing domains, particularly in those areas related to physical modelling.  
It drives the need for approximations both across and within physical time and length scales, leading to a  zoo of often interchangeable methods.
Beyond methodology, cost also drives the development of different hardware platforms and software frameworks, each offering a reduction in time-to-solution for particular methods.
Finally, the cost of simulations, often requiring tens of nodes running for tens of hours, drives a desire to reuse calculations.

These orthogonal approaches for tackling cost in computational science leads researchers to face a  continuum of choices when considering how to answer a given problem. 
This  manifests as an ever growing number of computational workflows providing the same measurement. 
To complicate matters, the optimal point in this continuum shifts depending on the exact question a researcher is asking and the time or cost constraints they are under.  

The goal of a workflow framework for computational workflows, particularly those that can be thought of as virtual-experiments, should be to help researchers to obtain measurements of properties they want, under the constraints they have, by leveraging the large space of choices available. 
We suggest that achieving this requires accounting for the prevalent use of approximation in the computational science domain as a basic design principle. 
Current workflow frameworks, which focus on helping developers to automate a well-defined task set in particular environments e.g. High Performance Computing (HPC), Cloud,  are not designed with this in mind.
In this article we:
\begin{itemize}
    \item Identify key operations that an approximation-aware computational workflow framework must support - sub-graph equivalence and sub-graph substitution - and describe how supporting these impacts key components of workflow frameworks, namely the description language, the knowledge-base, and the run-time. 
    \item Illustrate the potential of an approximation-aware workflow framework via a selection of initial prototypes and associated preliminary results.
    These include surrogate model support, automated workflow composition, and approximate memoization.
    \item Outline future directions of research and issues that we believe need to be addressed.
\end{itemize}

\section{Approximation in Computational Science}
\label{sec:approx-in-science}

Approximation is a key method in computational science for addressing cost pressures.  
For example in chemistry and physics, algorithms exist that use quantum mechanics to simulate systems. 
These usually deal with systems the size of Angstroms and time-scales of nanoseconds. 
At a certain stage the cost of modelling larger systems or longer times (100\AA, 100ns) requires moving to molecular dynamics (MD) which approximates the contribution of the electrons.  
When the cost of MD becomes a concern, researchers may employ methods like dissipative particle dynamics ($\mu$m, $\mu$s) that drastically simplify the entities considered (representing molecules as single particles) and so on. 

Within each scale, cost drives the continual development of new or improved methods.
Quantum chemistry serves as an exemplar. 
Density functional theory (DFT) was developed as an approximation for expensive Hartree-Fock (HF) methods. 
Then, more complex DFT methods were developed to account for physical phenomena that the original DFT approximation had omitted. 
Even within both HF and DFT methods multiple levels of theory can be applied, each presenting a different trade-off between cost and fidelity.

With the emergence of AI techniques, many research groups have explored creating AI-surrogate models to physical models  adding further to the zoo of methodologies \cite{carleo2019machine, butler2018machine}.
Such models promise a magnitude decrease in time-to-solution while matching physical model accuracy in many cases \cite{smith2017ani}. 
This leads to "surrogate" versions of physics-based computational workflows, where viable AI surrogate models are used in place of some (or all) of the physical ones.

Substituting a physical-model for an AI surrogate is a form of approximate computing, a domain with a rich history and academic literature. 
In approximate computing the aim is to create a approximate version of a calculation (surrogate) and dynamically substitute it, where appropriate, in place of the exact or "golden" calculation.
Typically, approximate computing deals with functions within a single HPC application, replacing them with optimized variants at the cost of accuracy~\cite{parasyris2021hpac, vassiliadis2015significance, vassiliad2016dco, laurenzano2016input, menon2018adapt, sharif2019approxhpvm}. 
In general, it aims to address three major technical challenges: a) identifying calculations for which to create surrogates~\cite{vassiliad2016dco, menon2018adapt}; b) generating well-performing surrogates~\cite{parasyris2021hpac, sharif2019approxhpvm}; and c) deciding when to replace a calculation with one of its surrogates (adjudication)~\cite{vassiliadis2015significance, laurenzano2016input}. 
The surrogate methods may be also be supported by, or implemented on, specialized hardware accelerators. 
For example, Google's Tensor Processing Unit (TPU)~\cite{jouppi2017datacenter} uses quantization and reduced precision tensor operations to accelerate machine learning workloads. 

\section{Related Work}
\label{sec:related-work}

There are a large number of workflow frameworks, each providing features targeting a specific domain (e.g. machine learning, scientific computations, etc) and/or certain platforms e.g. classic HPC stacks or Cloud environments like Kubernetes~\cite{burns2016borg}. 
For scientific workflows, some examples are AiiDA~\cite{huber2020aiida} and Pegasus~\cite{deelman2015pegasus} which are popular options for several scientific fields such as physics, astronomy, and bioinformatics.
MaestroWF~\cite{di2017maestro} is another example, targeted at HPC systems with integration for the Flux batch scheduler ~\cite{ahn2020flux} and support for reproducible workflow executions. 
There are many cloud-native frameworks for executing machine learning frameworks and deploying continuous-integration/continuous-deployment (CI/CD) pipelines.
Two of the most popular options are Argo~\cite{argo} and Tekton~\cite{tekton} which differentiate via their deep integration with Kubernetes. 
Ray~\cite{rayframework, rayworkflows} contains features especially helpful for machine learning workloads and is optimized for compute intensive Python tasks.
Kubeflow~\cite{kubeflow} is another cloud-native framework that specializes in orchestrating machine learning workloads, adopting Argo or Tekton as its underlying run-time system.

The common focus of these frameworks is providing assistance for automating task-sets on particular platforms and managing their output. 
The only approximate computing-like feature that some of these frameworks support is memoization, which uses a cache of task executions to eliminate redundant executions of equivalent tasks.
This can involve finding a previously executed instance of the exact same task (standard memoization~\cite{michie1968memo}).
However, researchers may be also satisfied with the output of a different method that provides the same, or similar, end result (approximate memoization~\cite{tziantzioulis2018temporal}). 
In either case it involves two basic operations: a) determining equivalence between tasks, and b) substituting the output of a task with the output of another.

With respect to the equivalence, it is desirable for a memoization method to identify as many redundant tasks (true positives) as possible; without incorrectly reusing cached results (false positives) or missing optimization opportunities (false negatives). 
We term this \textit{high-fidelity} and note that it is determined by the gap, if any, between the desired equivalence intent i.e. \textit{when} the system wants two tasks to evaluate as the same, and \textit{how} the system concretely performs this comparison (the criteria)~\cite{ieeescc2022memopaper}. 
Two other desirable characteristics of a memoization system are that it is \textit{fast}: memoizing must be faster than executing the tasks; and \textit{transparent}: memoization should be leveraged automatically and should not require the modification of workflows, or adoption of specific languages. 
Across the cited workflow frameworks a variety of memoization methods have been implemented. 
Some sacrifice transparency to be fast with high fidelity~\cite{flyte, babuji2019parsl, richard2018developing, heidsieck2019adaptive, pradal2008openalea, bavoil2005vistrails, altintas2004kepler, argo}. 
Others, are fast and transparent, but have fidelity issues with workflow nodes that consume file directories or digital artifacts (e.g. S3 buckets)~\cite{sevenbridges, kubeflow} resulting in false negatives or worse false positives. 

\section{Approximate Computing for Scientific Workflows}
\label{sect:approximation-aware-wf}

\begin{figure}
    \centering
    \includegraphics[width=\linewidth]{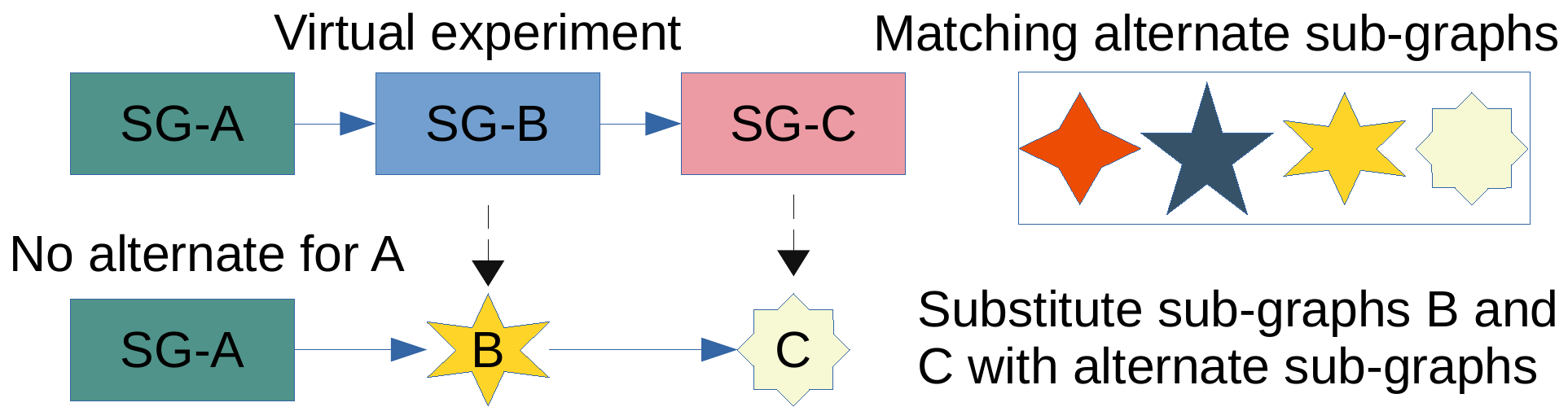}
    \caption{Illustration of the basic operations of equivalence testing and substitution in the context of virtual experiments. In this example, an initial virtual-experiment consists of 3 sub-graphs (motifs) and there is a pool of alternates determined by sub-graph equivalence. The sub-graphs and the alternates are evaluated with respect to a set of objectives. This result causes the workflow-runtime to perform two substitutions replacing the original sub-graphs B and C with alternates. The result is a new virtual-experiment which performs the same calculation but has different characteristics w.r.t. the objectives.}
    \label{fig:approximate-workflows}
\end{figure}

Approximate computing is based on three operations: a) determining calculation equivalence; b) applying substitution policies; and c) performing substitution. 
Our hypothesis is that the cost pressures which shape computational science - demanding faster, more approximate methods, reusing previously calculated results, and using different platforms - would be best handled by a workflow framework which is fundamentally designed to support approximate computing-like features. 
We term such a workflow framework \textit{approximation-aware}.

In \cite{garijo2014common} the authors examined 260 computational workflows and discovered that workflows generally evolve organically by wiring together smaller workflows often termed ``blocks" or ``motifs".
These smaller workflows are essentially sub-graphs that perform specific tasks and may appear in multiple larger workflows. 
Therefore an \textit{approximation-aware} workflow framework must provide the three approximation operations in the context of sub-graphs.
Concretely, determining sub-graph equivalence, calculating properties of sub-graphs for policy application,  and sub-graph substitution.
Figure \ref{fig:approximate-workflows} presents an example of the basic principles of equivalence and substitution.

Realising this requires system support at three levels - the workflow description language, the workflow knowledge-base, and the workflow run-time.
The workflow description must support sub-graph equivalence calculations; the workflow knowledge-base must support calculating, storing \& accessing sub-graph properties and relationships; and the run-time must support substitutions of sub-graphs in a workflow.

Combined, these features will lead to workflow frameworks that can handle large sets of computational workflows where each workflow could potentially satisfy a given measurement.
In this section we describe these components in more detail.
Then in Section \ref{sec:examples} we give concrete examples of how they would be leveraged to provide various approximate-computing like capabilities. 

\subsection{Sub-Graph Equivalence}

\begin{figure}[tb]
    \centering
        \subfloat[\centering]{\includegraphics[height=3.5cm]{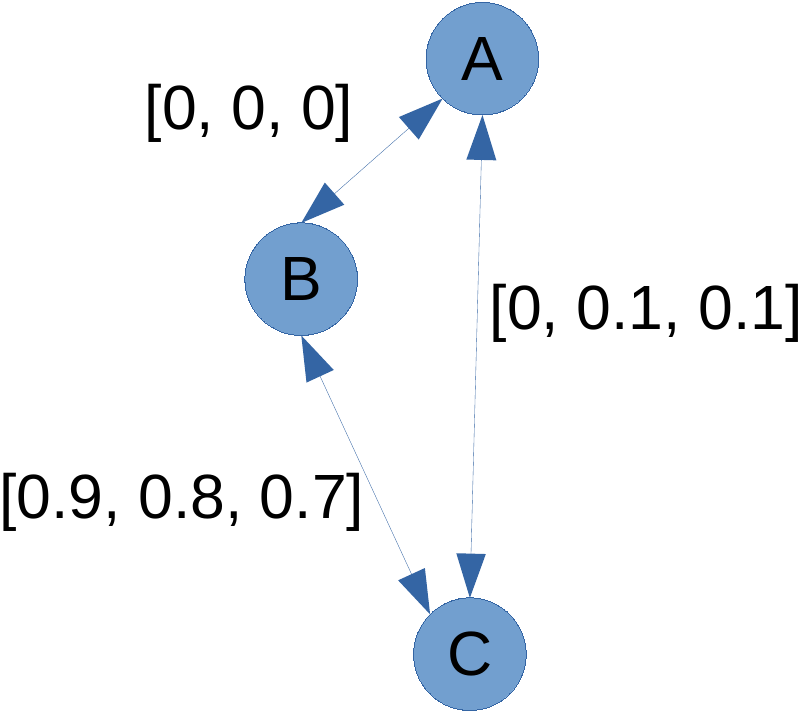}} \qquad{} \qquad{} \qquad{}
        \subfloat[\centering]{\includegraphics[height=3.5cm]{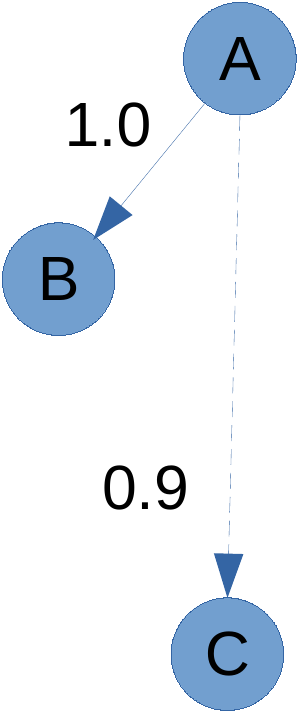}}
    \caption{Each node represents a sub-graph. On the left, edges indicate $[domain (d_{i}), function (f_{i}), co-domain (c_{i})]$ similarities. On the right, edges indicate known producer/consumer relationships (solid line) and hypothesized relationships (dotted line). For example, on the left we can see that $B$ consumes similar inputs to $C$ ($d_{i} = 0.9$) but not with A ($d_{i} =0.0$). On the right, $B$ is known to consume $A$'s outputs. Because the co-domain similarity of $B$ and $C$ is $0.9$ then we can hypothesize that $C$ can consume the outputs of $A$. Also, for $B$ and $C$ the function ($f_{i} = 0.8$) and co-domain ($c_{i} = 0.7$) similarity are both high. Therefore, we could potentially substitute $B$ for $C$ in workflows where $B$ consumes the output of $A$.}
    \label{fig:workflow-composition}
\end{figure}

When one method approximates another this means they satisfy a particular equivalence criteria. 
Hence determining the equivalence of two sub-graphs is the core mathematical operation that an approximation-aware workflow framework must support.  
We can divide this task into two parts: providing meaningful inputs to equivalence operations, and storing \& accessing the results of equivalence operations. 

\subsubsection{Supporting Sub-graph Equivalence Operations}
\label{sect:language-support}

We want to facilitate deciding if two workflow sub-graphs, $A$ and $B$, perform equivalent tasks - they take the same inputs and return the required outputs. 
This task is greatly simplified if the sub-graphs are not contaminated by non-functional and platform specific steps and details e.g. data-copying.
This means that the workflow description language must promote a high-level, \textit{abstractable}, description of the computational process. 
Here \textit{abstractable} indicates that the description itself can include platform and implementation specific details.
However, a platform and implementation independent view can be easily generated from it.

One way of realising this is through a combination of language features/syntax supported by run-time elements - we call this approach \textit{simple specification, smart-run-time}.
This approach seeks to offload many non-functional, platform specific, details to the run-time, which might otherwise appear as nodes in the workflow graph. 
This can include restarting, storage provisioning and data-copying operations among others. 
By removing them, the workflow sub-graph becomes simpler and better reflects the task it performs.

On the workflow description side, obtaining an abstract view is facilitated by cleanly separating the functional aspects of a node in the sub-graph - its inputs, task description, and steps - from non-functional details e.g. scheduler options. 
This allows these details to be easily omitted when a sub-graph is used for equivalence. 
Similarly, they can be easily added to the description to facilitate deployment on different platforms. 
It not only makes the comparison simpler, but also makes it straightforward to concretize a sub-graph for a particular platform once equivalence has been determined. 

\subsubsection{Sub-Graph Equivalence Operations}
\label{sec:equivalence-operations}

There are three equivalence operations to consider - see Figure \ref{fig:workflow-composition}
\begin{itemize}
\item Co-domain similarity ($c_{i}$): Do blocks produce the same data
\item Domain similarity ($d_{i}$): Do blocks consume the same data 
\item Function similarity ($f_{i}$): Do blocks do the same thing 
\end{itemize}
Note that for each operation multiple equivalence methods can be defined. 

The above operations also support answering the question “can blocks $A$ and $B$ be composed?”. 
This question  reduces to either
\begin{itemize}
  \item  Does $A$ produce the same data as the producers of $B$ (\textit{is $A$ like producers of $B$}). We call this the ``upstream" method.
  \item  Does $B$ consume the same data as the consumers of $A$  (\textit{is $B$ like consumers of $A$}). We call this the ``downstream" method.
\end{itemize}
Composition is important when performing substitution - see Section \ref{sec:substitution}

Note that depending on the similarity metric, you may not have to visit the producers directly to apply it.  
Given two nodes $A$ and $B$, where $B$ is consumer of $A$, information on the co-domain of $A$ can be built from the references to it in $B$. 
This in turn enables measuring co-domain similarity to $A$ by only accessing $B$.
 
\subsubsection{Storing Sub-Graph Equivalence Results}
\label{sect:subgraph-registry}

Given a large set of computational workflows, the number of equivalence relationships that could be computed will likely be very large.
As a result, they cannot be computed on the fly whenever the workflow framework needs to decide if a substitution could be performed. 
A solution is that a workflow knowledge-base actively performs and stores relationships and equivalence measurements. 

One approach is for the knowledge-base to maintain a graph, stored in a graph-database or similar, where each node represents a workflow sub-graph. 
Theoretically the graph will be fully-connected although many edges will have weight $0$.
The nodes in this graph can represent a workflow's sub-graphs in many ways. 
For example, a node can be associated with one or more hashes or feature-vectors (representations) which encode features of the sub-graph they represent; the node could include the sub-graph explicitly; or the node can include a reference to the source of the graph.

This graph will likely have at least three types of edges (see Figure \ref{fig:workflow-composition}).
The first type are equivalence edges (Section \ref{sec:equivalence-operations}).
These edges have weights which are the value of one or more similarity metrics: domain \& co-domain similarities (to aid in determining composability); functional similarities (for sub-graph matching).
The second edge-type is the producer-consumer edge. 
For a known producer/consumers pair the edge weight will be $1.0$.
For hypothesized connection the edge weight(s) are values of metrics which describe the likelihood that the blocks can be composed.
The final edge would denote ``sub-graph of" relationships. 

In addition to information on the sub-graph relationships, the workflow knowledge base should also provide access to computational workflow metadata and data, such as performance, accuracy, produced outputs, etc.
This functionality is already present to some extent in existing workflow frameworks e.g. Aiida and Kubeflow.  
In essence the knowledge-base should record as much as possible of information that is non trivial to extract.
This is key to enabling powerful systems that leverage these sources of information to make intelligent decisions (Section~\ref{sec:examples}).
For example, the information could be used to inform policy decisions (section \ref{sec:policy}) as well as to refine similarity metrics or create new ones.

\subsection{Sub-Graph Substitution}
\label{sec:substitution}

\subsubsection{Applying Sub-Graph Substitutions}
\label{sect:patching}

\begin{figure}[tb]
    \centering
    \includegraphics[width=\linewidth]{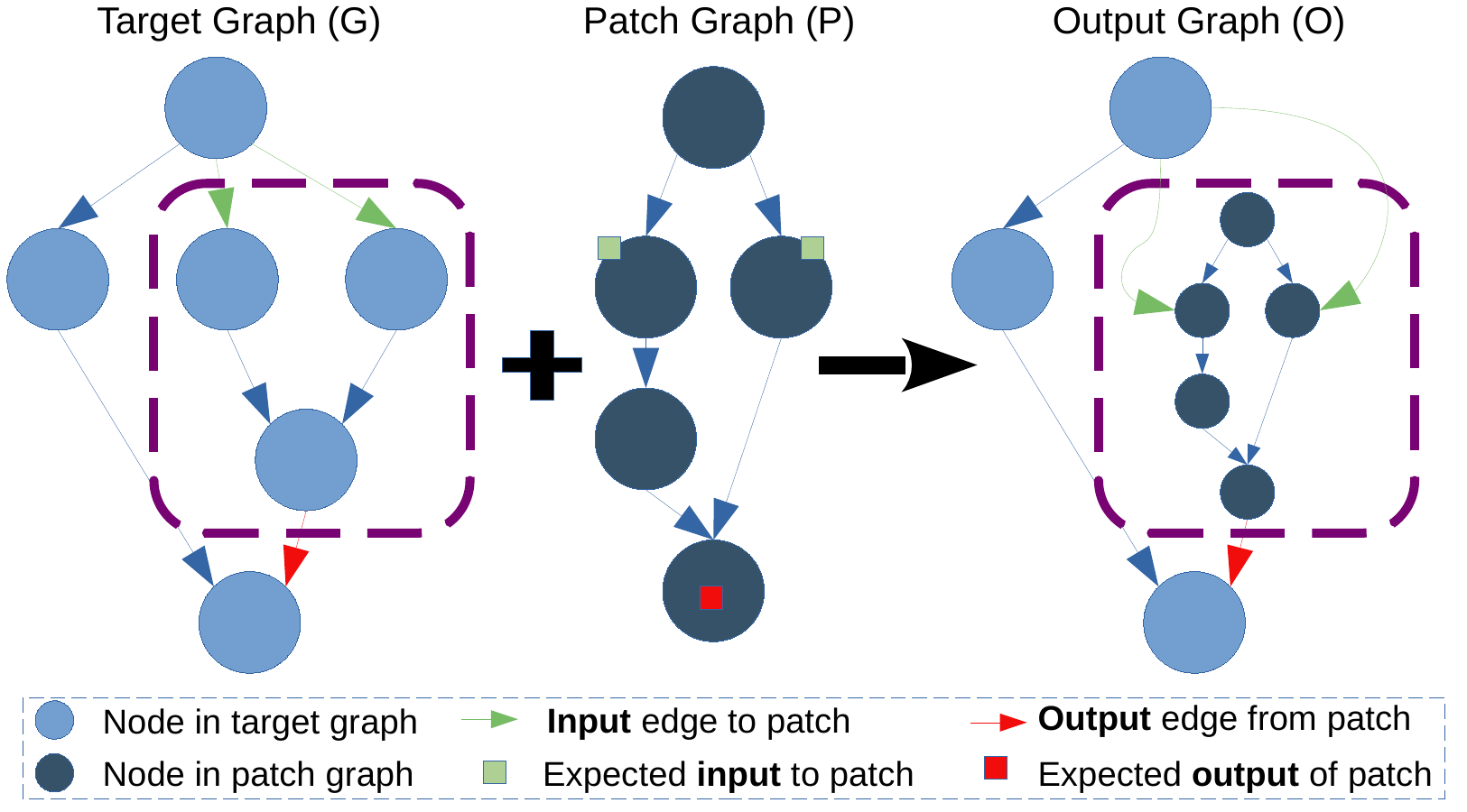}
    \caption{An example of patching a Target graph $G$ with a Patch graph $P$ to substitute a sub-graph of $G$. $P$ expects two inputs, from splice-points of $G$, and generates one output, that splice-points of $G$ can consume.
    The result is the Output graph $O$ on the right.
    Figure~\ref{fig:workflow-patches}.c) shows the splice-points at the top of $O$ providing inputs to $P$ and the splice-point at the bottom of $O$ consuming the output of $P$. 
     }
    \label{fig:workflow-patches}
\end{figure}

We consider that a sub-graph consists of one or more workflow nodes that may be connected with zero or more edges.
Each workflow node is associated with configuration metadata (e.g. definition of task, backend configuration, etc) and data (e.g. inputs/outputs, configuration files, etc).
The first is resident in the workflow-description while the second is external to the description.
When making a sub-graph substitution there are three main steps: a) creating/extracting a patch; b) identifying the splice points; and c) applying the patch.
A patch contains the sub-graph description, the data files that the nodes in the sub-graph require, along with other metadata required for the substitution e.g. schema of the inputs/outputs of the patch graph, removal/modification instructions for nodes in target graph, etc.

Before a patch $P$ can be applied to a target graph $G$ the workflow framework must first identify the splice points in $G$.
Splice points are the nodes and edges that either provide input to nodes in $P$ or consume their outputs.
Identifying the splice points requires single-node sub-graph equivalence tests between the original producers/consumers of $P$ and the nodes in $G$.
These can use input/output (domain/co-domain) or functional equivalence tests or a combination.
We note that the interface of $P$ is not guaranteed to be identical to the sub-graph of $G$ that will be substituted.
For example, the nodes in $P$ may require a different number of inputs.

Applying the patch then becomes a series of sub-graph substitutions on $G$.
Each sub-graph substitution aims to modify the current state of $G$ into a desired state. 
A sub-graph substitution involves two kinds of operations a) removing/updating zero or more nodes/edges in $G$, and b) inserting zero or more $P$ nodes into $G$.
Notably, there is a chance that after applying the sub-graph substitution operations the state of $G$ differs from the desired state.
For example, a splice point in $G$ that reads the output of node $N$ of $P$ may expect $N$ to have a specific command-line argument.
If $N$ does not fulfill this expectation, the workflow framework flags this inconsistency as a ``conflict" and attempts to resolve it.
The difficulty of this operation depends strongly on the level of annotation supported by the workflow description language e.g. does it allow labelling of optional command line arguments.

\subsubsection{Determining Sub-Graph Substitutions}
\label{sec:policy}

Automatically determining which sub-graphs are substitutable by which other sub-graph is a challenging problem, particularly as computational workflows become large.
There are 3 steps for creating a sub-graph substitution policy: a) specifying the parameters/objectives; b) exploring the space of potential substitutions using one or more agents; c) crafting the final substitution policy using a superintendent agent.
The first step involves defining a cost-function to evaluate the different substitutions.
For example, this may leverage some domain-specific language, or even process natural language text to extract the parameters/objectives.
The next step potentially involves multiple agents, each of which suggests a sub-graph substitution plan for the entire virtual-experiment.
For example, each of these agents can specialize in a particular task (e.g. optimize performance, optimize accuracy, optimize cost, etc).
An agent produces a substitution plan which contains one or more sub-graph substitutions as well as metadata about the expected benefits (e.g. performance, accuracy, etc) and costs (e.g. downtime, performance, accuracy, etc) of the plan.
In the final step, a superintendent agent produces the final sub-graph substitution policy.
There are many different ways to implement such a superintendent.
The most straightforward solution is to inspect all the proposed sub-graph substitutions, and pick the one that has the lowest value of the cost-function.
Alternatively, the superintendent could pick and choose different parts of the many sub-graph substitutions that the first level agents suggested to see if a lower cost-function value could be obtained.

\section{Prototypes and Initial Evaluations} 
\label{sec:examples}

For a number of years, ourselves and our collaborators have focused on computational workflows for materials design \cite{mcdonagh2021can, abdelbaky2014exploring, johnston2016, johnston2020, conchuir2020, klebes2020}.
This involved designing and running large scale computational chemistry workflows on machines from BG/Q\textsuperscript{TM}, to IBM POWER8\textsuperscript{TM} through to clusters managed by Kubernetes~\cite{burns2016borg}. 
The frequent changing of hardware-architecture, data-architecture, and schedulers, led us to abstract these details from the workflow description and handle them instead in the run-time (Section~\ref{sect:language-support}).
For example, we handled scheduling via pluggable backends, abstracted storage and data-movement from task definitions and introduced a variable layering scheme to allow easy customization of workflows for different environments. 
Automation and robustness was critical as we were running iterative calculations requiring up to 1600 IBM POWER8\textsuperscript{TM} cores and 1 week of compute time with hundreds of independent co-processing tasks. 
This led us to also hand-off restart details from the workflow description i.e. each task in the description represents a logical step, rather than a concretized task execution.
These workflow-runtime and description language features gave us the basis to support sub-graph equivalence operations (Section \ref{sect:language-support}).
Subsequently we've built, and are building, a number of prototypes on this foundation to explore different aspects of an approximation-aware workflow framework, some of which we describe here.  

\subsection{Approximate Memoization}
\label{sect:memoization}

When working on searching large discovery spaces for material candidates, we encountered high reuse of sub-graphs across multiple virtual-experiments. 
To handle this, we introduced a prototype approximate memoization scheme (Section \ref{sec:related-work}) for computational workflows which helped us accelerate the search \cite{ieeescc2022memopaper, photoresist, pyzer2022accelerating}.
We found this was facilitated by the abstractions we had introduced which made it straightforward to implement suitable single-node equivalence criteria.
This was based on the node's interface and the chain of nodes leading to it, specifically;
\begin{enumerate}
\item Any inputs which are not outputs of other nodes must be bit-wise identical.
\item For inputs coming from producer tasks, those tasks have identical interfaces.
\end{enumerate}
We could use the chain of tasks as we knew it would reflect the functional task-chain uncontaminated by instance-specific details like number of restarts. 
The equivalence criteria also did not use platform or execution-specific values. 
This enabled us to support memoization between tasks that executed on distinct and heterogeneous execution environments. 
We implemented a workflow knowledge-base for storing metadata on executed tasks (nodes).
This assigned identifiers in the form of hashes to tasks (Section \ref{sect:subgraph-registry}) that could be searched for  tasks that executed in the past.
The run-time performed the substitution by skipping the original task and writing the output of equivalent tasks to its output store.

In our study, we showed the equivalence criteria provided higher-fidelity compared to existing methods as it avoided classes of false negative/positives present in prior works. 
It was also fast, requiring (on average) $\mathcal{O}(ms)$ to generate the hash of a node in our experiments. 
This means that memoization can be worthwhile even for tasks which execute in $\mathcal{O}(sec)$. 
In our experiments, memoization offered speedup of up to $10.55\text{x}$ without any configuration/hints by workflow developers or users.

However, we took some shortcuts in the prototype that can be improved upon.
We skipped the true substitution of the node in the description. 
This could cause issues when inspecting the executed workflow description later, particularly if a less strict equivalence criteria was adopted in future.
In addition, the current prototype is restricted to using equivalence tests at the granularity of single nodes. 
Expanding this to leverage full sub-graph equivalence, and hence a memoization method that acts at a sub-graph level, would be powerful. 
For example, it would allow memoizing a N-node sub-graph using one quantum chemistry code with a M-node sub-graph that uses another quantum chemistry code, and that produces the same numerical analysis but has improved performance. 
Achieving this would require a more complete implementation of the approximation-aware knowledge-base functionality, as described in \ref{sect:subgraph-registry}, along with more advanced sub-graph equivalence methods.

\subsection{Surrogate Model Support}
\label{sect:surrogate-model-support}

In Section \ref{sec:approx-in-science} we mentioned recent research efforts in creating AI surrogates of physical models.
We are currently building capability to support the development and use of such AI surrogates in computational workflows. 
For clarity, in the following we distinguish between a surrogate model and a surrogate calculation. 
The first is the actual model (usually a function with some parameters) that replaces a physical model (also a function with some parameters).
The second is a sub-graph that uses the surrogate model that can be used to substitute a sub-graph using a physical model. 

In particular we want to address the adjudication challenge - when to substitute a surrogate for the physical calculation. 
Adjudication is crucial to both the performance of the overall computation as well as its correctness. 
It requires specific run-time features of an approximation-aware workflow framework: identification of surrogates for physical calculations; effective run-time surrogate selection policies; and dynamic substitution capabilities. 

In addition, adjudication support features like canary testing~\cite{laurenzano2016input} are highly desirable to reduce burden on developers. 
This involves automatically, and transparently (in background), launching surrogate-calculations alongside matching physical calculations, to build a data-set for assessing accuracy (canary testing). 
This also leverages the key features of an approximation-aware workflow stack.

For initial versions of this prototype we will adopt explicit indication of physical/surrogate relationships.  
This requires manual identification of equivalent sub-graphs and associated information for patching e.g. splice points (section \ref{sect:patching}).
Nevertheless, supporting it requires a more complex version of the workflow knowledge-base than what we used in our approximate memoization prototype. 

In section \ref{sec:policy} we outline a general protocol for deciding if a sub-graph should be substituted.
We term this a ``prior policy" as it is applied before substitution. 
AI-surrogates also bring additional complexity as they can potentially provide information on the uncertainty of their predictions w.r.t the physical model they were trained on. 
This can be a feature provided by the virtual-experiment before the calculation begins e.g. a model prediction, in which case it requires additional run-time and workflow-specification support. 
It could also be provided as part of the result that the surrogate model calculates. 
This opens the possibility of rolling back the application of an AI-surrogate model i.e. support ``posterior policies". 
This would require extra complexity in the workflow knowledge-base. 

\subsection{Automatic Computational Workflow Composition}
\label{sect:composition}

\begin{figure}[tb]
    \centering
    \includegraphics[width=\linewidth]{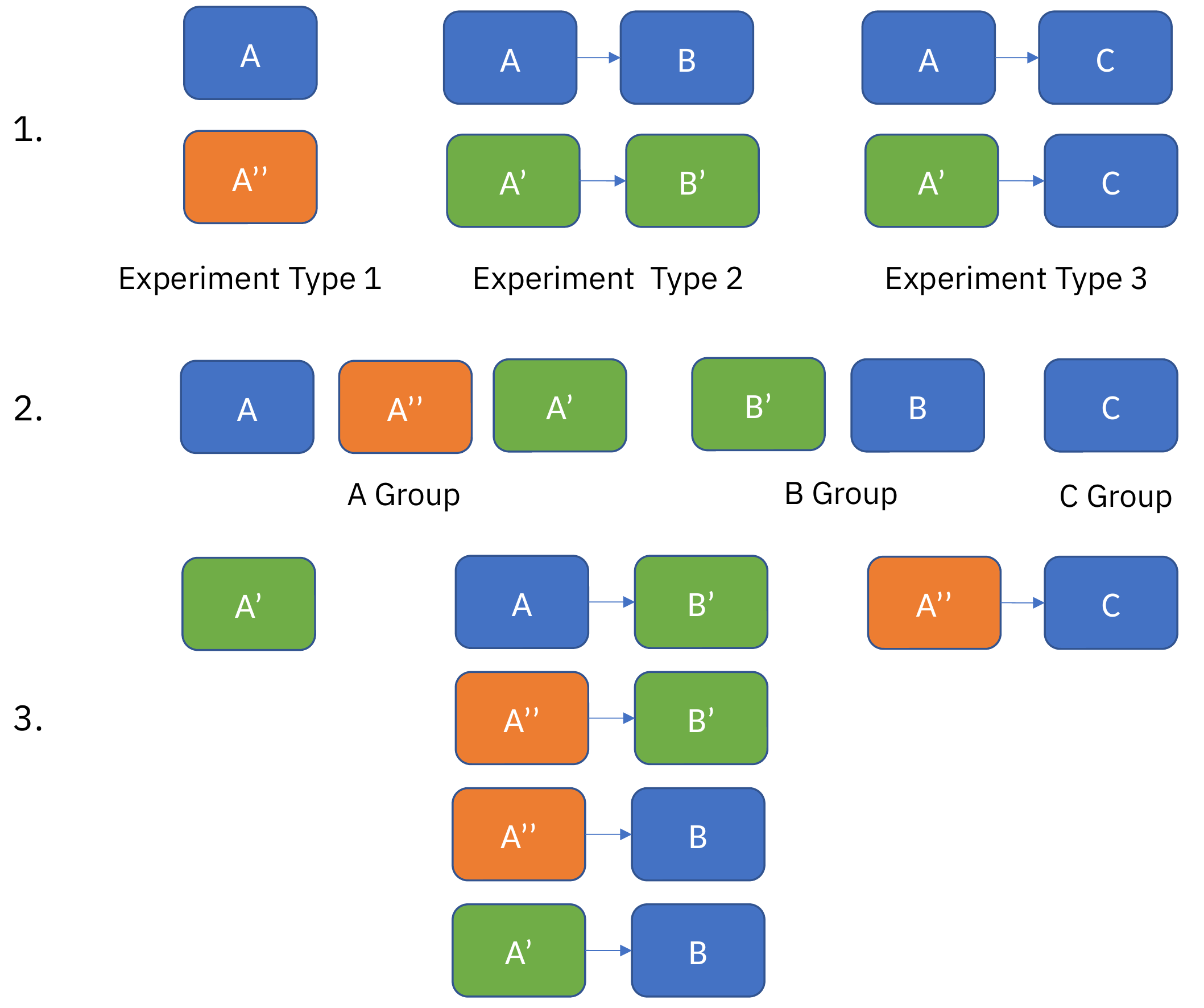}
    \caption{(1) We started with three distinct virtual-experiments each with two alternates. Two of the virtual-experiments were composed of two distinct sub-graphs or blocks. Our factoring algorithm was successfully able to identify the sub-graphs. (2) Using the Weisfeiler-Lehman graph isomorphism method we were able to recognize both the six distinct block types and the three block groups. (3) In combination with a sub-graph substitution capability the knowledge of the blocks and groups would allow one to create ten new virtual-experiments - six following the three templates shown in \ref{fig:composition}.1 and four single block virtual-experiments that were not previously exposed. }
    \label{fig:composition}
\end{figure}

As we have discussed, workflow developers may logically partition their workflow into sub-graphs which perform specific reusable tasks. 
However, it is useful to have a method to automatically decompose or factor workflows into such independent sub-graphs.  

Factoring a graph $W$ is equivalent to creating a quotient graph of $W$.
Creating a quotient graph requires defining an equivalence relationship which designates if two nodes $u$ and $v$ are in the same block. 
We experimented with a variety of equivalence relationships.
One we found most successful was the following:
\begin{equation}
\label{eq:decomposition}
E = \prod_{p \in P}  ( u \in p \land v \in p )
\end{equation}
where $P$ is all sub-graphs to leaf nodes of $W$.
Equation \ref{eq:decomposition} states that given a set of sub-graphs $P$, nodes $u$ and $v$ are equivalent if they appear together in all sub-graphs leading to a given leaf-node and never separately.


We trialed this method on the set of six virtual experiments we developed in \cite{pyzer2022accelerating}. 
There were three distinct experiments each with two alternates (see Figure \ref{fig:workflow-composition}.1).  
These experiments had organically evolved from each other and we knew four consisted of two distinct steps. 
We found equation \ref{eq:decomposition} successfully partitioned each workflow into its distinct functional blocks. 
\par
Although we knew blocks were similar, we sought to show this algorithmically. 
We trialed the Weisfeiler-Lehman graph isomorphism method \cite{shervashidzeSLMB11}, using edges labelled with the data consumed and nodes labelled with (a) the name of the node, and (b) the nodes command line. 
We found using the command-line successfully identified the six unique blocks and the name successfully grouped the blocks into the three groups (see Figure \ref{fig:workflow-composition}.2) .
\par
In combination with a sub-graph substitution capability (section \ref{sec:substitution}) this opens the possibility to recombine the blocks. 
In total, these blocks could be recombined into six new valid alternate virtual-experiments, following the original three templates (see Figure \ref{fig:workflow-composition}.3) .
In addition, each block itself is a valid computational workflow and hence the four blocks ($A', B, B', C$) could also be extracted.
Thus, from six initial experiments the combination of automatic factoring and composition can add a further ten.

\section{Discussion and Conclusion}

The examples in Section~\ref{sec:examples} give a flavor of the potential capabilities of an approximation-aware framework. 
However, these prototypes have a number of limitations.
Resolving these limitations requires deeper investigation into a variety of systemic and algorithmic topics. 
Such a framework fundamentally relies on equivalence operations at the granularity of workflow graphs.
In our prototypes, we adopted basic equivalence methods (Section~\ref{sect:memoization} and \ref{sect:composition}) or human-annotated equivalence (Section \ref{sect:surrogate-model-support}). 
In the future we would like to investigate different equivalence methods for the three equivalence operations (Section~\ref{sec:equivalence-operations}).
We also wish to evaluate how different method/operation type pairs perform for various tasks. 

Related to this, there are questions around sub-graph representations e.g. how can effective representations be generated?, are different representations better for different tasks?, and how to interpret non-binary equivalence measures e.g. what does it mean that one sub-graph is 90\% the same as other? 
Differences can be due to inherently different calculations but could also be syntax induced e.g. due to the way tasks are described there is uncertainty in how elements, like inputs, map to each other (Section \ref{sect:patching}). 

This issue illustrates the important role that the workflow specification plays in determining equivalence. 
In particular, there is a trade-off between relaxing the syntax to allow greater potential matching e.g.  by adopting weak-typing, and how certain equivalence methods rely on the information transmitted by the syntax.
We adopted a quite relaxed syntax in our prototypes but we can see how greater annotations would make equivalence tasks easier. 
Two lines of investigation here are using introspection and AI to provide annotations e.g. identifying types for outputs, and/or to relying on emergent convention/community driven efforts to remove friction. 
For example, an IDE could raise a flag when a specification is causing equivalence issues.
A developer could then add just enough extra information to resolve these issues.

A critical system component in the framework is the workflow knowledge-base. 
We noted in Section \ref{sect:subgraph-registry} that a knowledge-graph that is built to hold sub-graph similarity relationships will be densely connected.
Additionally, we consider that constructing such a component with the goal of it handling a real load is an open question.
Answering this question relies on knowledge on the frequency and type of queries, along with deciding where reasoning should occur, how workflow/workflow-instances are connected, and how supporting data-stores e.g. containing actual output files, are integrated. 
Nevertheless, although the size of the pool of computational workflows will raise technical issues for any workflow knowledge-base, the overall capability of the system can only improve as its computational-workflows corpus increases in size.

An approximation-aware workflow framework is one that understands that there are multiple ways to realize a given result (measurement), and that can support researchers in making the best choice. 
In particular we see this was having significant impact on problems involving search of large entity spaces where the usual aim is to find entities (systems) with desired characteristics under some cost constraint c.f. material discovery. 
The resulting cost-constrained optimisation often has a large number of potential methods i.e. computational workflows, to measure these key-characteristics. 

However, from the researcher's view the exact method adopted is often a means-to-an-end. 
Hence the ability to construct the possible approximations, then learn their characteristics, and finally select the best approximation is critical for an effective search. 
For example, fast methods could be chosen to quickly probe regions, and occasionally more detailed methods could confirm or calibrate the predictions of the fast methods.
Furthermore, the information gained on the methods and their relationships can be used to accelerate future searches. 
Hence, a workflow system that can natively support this task, allowing complex and intelligent capabilities to emerge out of a small set of basic features and principles, will be invaluable.

\begin{acks}
Parts of this work were supported by the STFC Hartree Centre's \textit{Innovation: Return on Research programme}, funded by the Department for Business, Energy \& Industrial Strategy.

\end{acks}

\bibliographystyle{ACM-Reference-Format}
\bibliography{bibliography}

\end{document}